\documentclass[11pt]{iopart}
\usepackage{graphicx}
\usepackage{mathbbol}
\makeatletter
\AtBeginDocument{\@ifpackageloaded{natbib}{\ifNAT@numbers\if@filesw\immediate\write\@auxout{\string\global\string\NAT@numberstrue}\fi\fi}{}}
\makeatother
\begin{document}

\title[Mathematical understanding of violation of detailed balance condition]{Mathematical understanding of   detailed balance condition violation and its application to Langevin dynamics}
\author{Masayuki Ohzeki and Akihisa Ichiki}
\address{Department of Systems Science, Graduate School of Informatics, Kyoto University, 36-1 Yoshida Hon-machi, Sakyo-ku, Kyoto 606-8501, Japan}
\address{Green Mobility Collaborative Research Center, Nagoya University, Fro-cho, Nagoya, 464-8603, Japan}

\begin{abstract}
We develop an efficient sampling method by simulating  Langevin dynamics with an artificial force rather than a natural force by using the gradient of the potential energy.
The standard technique for sampling following the predetermined distribution such as the Gibbs-Boltzmann one is performed under the detailed balance condition.
In the present study, we propose a modified Langevin dynamics violating the detailed balance condition on the transition-probability formulation.
We confirm that the numerical implementation of the proposed method actually demonstrates two  major beneficial improvements: acceleration of the relaxation to the predetermined distribution and reduction of the correlation time between two different realizations in the steady state.
\end{abstract}
\ead{mohzeki@i.kyoto-u.ac.jp}
\submitto{J. Conf. Ser:}
\maketitle
\normalsize

\section{Introduction}
To investigate the statistical properties of a many-body interacting system, often sampling following the predetermined distribution is employed.
Because of strong correlation between degrees of freedom, it is usually intractable to compute the expectation, covariance, etc. following the distribution by  direct manipulation.
Instead, we often use a sampling method.
One of the well-known techniques is the Markov-chain Monte Carlo (MCMC) method, which imitates the stochastic dynamics of the discrete-time master equation.
This is a promising method used to evaluate the desired quantities, but it sometimes takes a long time to convert the initially prepared system into a suitable state for computation  because the method imitates the relaxation of the stochastic dynamics to the predetermined distribution as a steady state.
Innovative studies to enhance its sampling efficacy   and accelerate the relaxation to the predetermined distribution have been extensively reported \cite{Swendsen1987,Hukushima1996,Neal2001,Ohzeki2010a,Ohzeki2010b,Ohzeki2011proc,Ohzeki2011proc2,Ohzeki2012d}.

Basically, we construct a transition matrix to relax the initially prepared system to the steady state, which corresponds to the predetermined distribution in the MCMC method.
A series of  instantaneous realizations is generated by simulating the multiplication of the transition matrix.
After sufficient  iteration steps, the obtained series of  realizations follows the predetermined distribution.
This is assured by the balance condition (BC) on the transition matrix. 
In general, it is difficult to solve the BC to compute the transition matrix.
The standard approach for finding the solution is by use of the detailed balance condition (DBC). 
Because the DBC always satisfies the BC, we may instead solve the DBC for construction of the transition matrix.
However, the solution under the DBC is not necessarily an optimal choice of the transition matrix.
Recent studies reveal the superior performance, which is confirmed by means of the faster convergence to the predetermined distribution, by using   nontrivial solution neglecting the DBC \cite{Suwa2010,Turitsyn2011,Fernandes2011}.
As detailed also in the present paper, the present authors provided its mathematical validation through an analysis on the eigenvalues of the transition matrix in a previous study \cite{Ichiki2013}.

The method for sampling is not limited to the MCMC method.
The simplest technique is to implement  Langevin dynamics, which is available for the stochastic dynamics of continuous variables in continuous-time evolution.
In this alternative, does the analogous procedure for achieving faster convergence to the predetermined distribution exist?
The time evolution of the probabilistic distribution for the degrees of freedom driven by  Langevin dynamics is written by using the Fokker-Planck equation.
This is the continuous equation for the instantaneous distribution function.
In the steady state, the divergence of the flow of the distribution must vanish.
This is analogous to the BC in the context of the master equation.
The condition that the steady flow itself vanishes corresponds to the DBC.
We often demand that the force, which drives the system to simulate the natural dynamics, be the gradient of the potential energy.
This natural force holds the analogous condition to the BC and can be confirmed to satisfy the DBC in a different formulation.
However, it is not necessarily the case if we desire to reach the predetermined distribution as fast as possible.
We have  room to design a force to speed up the convergence to the desired distribution.
In the present study, we propose several forces to violate the condition analogous to the DBC while satisfying the analogous condition to the BC in the steady state.

The rest of the paper consists of the following sections.
We review several ways to construct the transition matrix  neglecting the DBC in the MCMC method in the next section.
After that, we introduce the Langevin equation and the corresponding Fokker-Planck equation in the third section.
In this section, we find several nontrivial solutions violating the analogous condition to the DBC.
The following section presents several numerical tests of the artificial forces to speed up convergence to the desired distribution.
Before closing the present paper, we show the mathematical assurance of the faster convergence by analyzing the eigenvalues of the operators in the Fokker-Planck equation.
In the last section, we summarize our study and provided direction for future plans.

\section{Sampling method: Monte Carlo simulation}
Let us briefly review the method to achieve faster convergence to the desired distribution in an MCMC simulation from its fundamental ingredients \cite{Nishimori2011}.
\subsection{Master equation}
Markov-chain Monte Carlo simulations imitate the discrete-time evolution of the master equation given by
\begin{equation}
P({\bf y},t+1) = \sum_{{\bf x}}P({\bf y}|{\bf x})P({\bf x},t),
\end{equation}
where $t$ is a discrete time and $P({\bf x},t)$ is the instantaneous distribution for the degrees of freedom ${\bf x}$.
The transition matrix $P({\bf y}|{\bf x})$ defines the rule of the stochastic dynamics in the master equation.
At least, the conservation of the probability (CP) demands that the transition matrix should satisfy
\begin{equation}
\sum_{{\bf y}}P({\bf y}|{\bf x}) =1.
\end{equation}
In addition, the nondiagonal elements of the transition matrix must be non-negative.
\subsection{Balanced condition and its solution}
In MCMC simulations, we numerically perform an iterative update of the instantaneous distribution following the master equation.
The distribution function in the steady state $P^{\rm (ss)}({\bf x})$ is defined by the balance condition 
\begin{equation}
P^{\rm (ss)}({\bf y}) = \sum_{{\bf x}}P({\bf y}|{\bf x})P^{\rm (ss)}({\bf x}).\label{BC}
\end{equation}
By use of the transition matrix satisfying the BC, we obtain the desired distribution function, which is set to be $P^{\rm (ss)}({\bf x})$, after sufficient  iteration steps.
For simplicity, we restrict ourselves to the case  aimed at generating the Gibbs-Boltzmann distribution $P^{\rm (ss)}({\bf x}) = \exp(-\beta E({\bf x}) + \beta F)$, where $F$ is the free energy and $\beta$ is the inverse temperature.

It is not simple to obtain the nontrivial solution under the BC.
The most standard way to find the solution is to use the detailed balance condition 
\begin{equation}
P({\bf x}|{\bf y})P^{\rm (eq)}({\bf y}) = P({\bf y}|{\bf x})P^{\rm (eq)}({\bf x})
\end{equation}
for ${\bf y} \neq {\bf x}$.
One can confirm that the DBC indeed satisfies the BC by use of CP.
However, this is not a unique way to find the nontrivial solution under the BC.
Recent studies on  DBC violation have led to  new way to construct the transition matrix, as introduced below.
 
\subsection{Violation of the detailed balance condition}
The construction of the transition matrix can be roughly categorized as
1. optimization following some rule of the transition matrix neglecting the DBC and
 construction of a duplicate system. 
\subsubsection{Suwa-Todo method}
The former type of  method was recently proposed by Suwa and Todo \cite{Suwa2010}.
They considered the minimization of the rejection rate, which corresponds to the diagonal elements of the transition matrix.
In numerical implementation of the MCMC method, we generate a candidate as the next configuration at each update.
Then we accept the candidate following the probability determined by the nondiagonal elements.
The average of the rejection rate is thus given by the summation over the diagonal elements of the transition matrix. 
The diagonal elements are given by the summation of the nondiagonal elements from CP.
To eliminate the diagonal elements, the Suwa-Todo method thus allocates the nondiagonal elements while neglecting the DBC but satisfying the BC and CP.
Let us rewrite the transition matrix into the probabilistic flow as
\begin{equation}
V({\bf y}|{\bf x}) = P({\bf y}|{\bf x})P^{\rm (ss)}({\bf x}).
\end{equation}
Then the BC and CP are given as 
\begin{eqnarray}
\sum_{{\bf x}}V({\bf y}|{\bf x}) &=& P^{\rm (ss)}({\bf y}),\\
\sum_{{\bf y}}V({\bf y}|{\bf x}) &=& P^{\rm (ss)}({\bf x})
,\end{eqnarray}
respectively.
The symmetry of the probabilistic flow $V({\bf x}|{\bf y})=V({\bf y}|{\bf x})$ reads the DBC.
The Suwa-Todo method finds the rejection-free solution violating the symmetry of the probabilistic flow.
We do not follow their method in the present paper but we remark on its efficacy  from  theoretical aspects.
To eliminate the diagonal elements of the transition matrix, we may change the nondiagonal elements of the transition matrix.
As also shown below in a different formulation, the asymmetry in the transition matrix can accelerate  relaxation to the steady state, as shown by the present authors in Reference \cite{Ichiki2013}.
We thus conclude that  part of the efficacy of the Suwa-Todo method comes from the asymmetry in the transition matrix, which violates the DBC.
However, Suwa and Todo simultaneously reduce the rejection rate by decreasing the diagonal part.
In the present paper, we restrict ourselves to the case in which the diagonal elements are fixed.
In this sense, the efficacy of the Suwa-Todo method is not yet completely understood.

\subsubsection{Skewed detailed balance condition}
The latter type is the skewed detailed balance condition (SDBC) \cite{Turitsyn2011}.
We prepare a duplicate system by introducing a replica of the original system.
For each system, we construct  transition matrices $P_1({\bf x}|{\bf y})$ and $P_2({\bf x}|{\bf y})$.
Then we define the following master equation:
\begin{equation}
P_i({\bf y},t+1) = \sum_{{\bf x}}P_i({\bf y}|{\bf x})P_i({\bf x},t) + Q_i({\bf y})P_{\bar{i}}({\bf y}) - Q_{\bar{i}}({\bf y})P_i({\bf y}),
\end{equation}
where $P_i({\bf x},t)$ is an instantaneous distribution for each system.
The swap transition from $i$ to $\bar{i}$ (the counter system of $i$) is given by $Q_i({\bf y})$ and $Q_{\bar{i}}({\bf y})$ for the inverse manner.
The balanced condition for the duplicate system holds as 
\begin{equation}
P^{\rm (ss)}({\bf y})= \sum_{{\bf x}}P_i({\bf y}|{\bf x})P^{\rm (ss)}({\bf x}) + Q_i({\bf y})P^{\rm (ss)}({\bf y}) - Q_{\bar{i}}({\bf y})P^{\rm (ss)}({\bf y}).
\end{equation}
We here impose that the steady state shares the same distribution as $P^{\rm (ss)}({\bf x})$.
One of the generic solutions satisfying the BC is known as the SDBC.
The transition matrix under the SDBC is given by
\begin{equation}
P_i({\bf y}|{\bf x})P^{\rm (ss)}({\bf x}) = P_{\bar{i}}({\bf x}|{\bf y})P^{\rm (ss)}({\bf y})
\end{equation}
for ${\bf y} \neq {\bf x}$ and the transition between each system should simultaneously satisfy
\begin{equation}
P_1({\bf y}|{\bf y}) - P_2({\bf y}|{\bf y}) + Q_1({\bf y}) - Q_2({\bf y}) = 0.
\end{equation}
In other words, the BC for each system in the ordinary form as in Equation (\ref{BC}) is violated.
Instead, we demand the BC for the duplicate system, in which the transition between each system is taken into account.

\subsection{Ichiki-Ohzeki asymmetric transition matrix}
The violation of the DBC can be recast as the asymmetry of the transition matrix.
We rescale the transition matrix and the instantaneous distribution function as
\begin{eqnarray}\nonumber
R({\bf x},t) &=& \frac{P({\bf x},t)}{\sqrt{P^{\rm (ss)}({\bf x})}} ,\\
W({\bf y}|{\bf x}) &=& \frac{1}{\sqrt{P^{\rm (ss)}({\bf y})}}P({\bf y}|{\bf x})\sqrt{P^{\rm (ss)}({\bf x})}.
\end{eqnarray}
Then the master equation can be rewritten as 
\begin{equation}
R({\bf y},t+1) = \sum_{\bf x}W({\bf y}|{\bf x}) R({\bf x},t)\label{PV}.
\end{equation}
Two of the conditions, namely, the BC and CP, that the master equation must satisfy are simply written as
\begin{eqnarray}
\sum_{\bf x}W({\bf y}|{\bf x}) \sqrt{P^{\rm (ss)}({\bf x})}&=& \sqrt{P^{\rm (ss)}({\bf y})} ,\\
\sum_{\bf y}\sqrt{P^{\rm (ss)}({\bf y})} W({\bf y}|{\bf x}) &=& \sqrt{P^{\rm (ss)}({\bf x})},
\end{eqnarray}
respectively.
In this expression, the DBC is interpreted as  symmetry of the rescaled transition matrix as $W({\bf y}|{\bf x}) = W({\bf x}|{\bf y})$.
The present authors have proven that the asymmetry of the rescaled transition matrix, violation of the DBC, leads to  faster acceleration of relaxation to the steady state then that driven by using the symmetric part of the rescaled transition matrix \cite{Ichiki2013}.

We emphasize that the essential aspect of speeding up the relaxation  is the asymmetry of the rescaled transition matrix. 
To introduce the asymmetry, for example in the SDBC, the transition to the duplicate system was considered \cite{Turitsyn2011}.
In contrast, in the Suwa-Todo method \cite{Suwa2010}, reducing the rejection rate and the nondiagonal elements were changed from those under the DBC as a result.

\section{Sampling method: Langevin equation}
\subsection{Langevin equation and its corresponding Fokker-Planck equation}
To generate the stochastic dynamics, instead of the master equation, we may use the overdamped Langevin dynamics defined as
\begin{equation}
d{\bf x} = {\bf A}({\bf x})dt + \sqrt{2T}d{\bf W},
\end{equation}
where ${\bf x}$ is the vector standing for the $N$-dimensional degrees of freedom and $d{\bf x}$ is the infinitesimal change of ${\bf x}$ during $dt$.
In addition, ${\bf A}({\bf x})$ is a time-independent force vector, $T$ is the temperature, and $d{\bf W}$ is the Wiener process for the $N$-dimensional degrees of freedom.
The force is not necessarily given by $-{\rm grad}E({\bf x})$ as in equilibrium.
The purpose is hereafter to find an artificial force to realize faster convergence to the desired distribution.

For achieving this goal, let us write the corresponding Fokker-Planck equation to the above Langevin equation:
\begin{equation}
\frac{\partial P({\bf x},t)}{\partial t} = - {\rm div}{\bf J}({\bf x},t),
\end{equation}
where ${\bf J}({\bf x},t)$ is the probabilistic flow defined as
\begin{equation}
{\bf J}({\bf x},t) = \left({\bf A}({\bf x}) - T {\rm grad}\right)P({\bf x},t).
\end{equation}

\subsection{Divergence-free condition}
When the system is in any steady state, the divergence of the probabilistic flow vanishes.
Therefore, we demand the following equality in the steady state:
\begin{equation}
0 = - {\rm div}{\bf J}^{\rm (ss)}({\bf x}) \label{BCLan},
\end{equation}
where
\begin{equation}
{\bf J}^{\rm (ss)}({\bf x}) = \left({\bf A}({\bf x}) - T {\rm grad}\right)P^{\rm (ss)}({\bf x}).
\end{equation}
We refer to the above equality as the divergence-free condition.
Our problem is to find a solution ${\bf J}^{\rm (ss)}({\bf x})$ satisfying the divergence-free condition.
The above equality, which holds in any steady state, is analogous to the BC in the context of the master equation.
Hereafter the important thing is to find a nontrivial solution under the divergence-free condition.
The solution is violating the DBC but under the BC in the same spirit as in the previous studies for acceleration of the MCMC method \cite{Suwa2010,Turitsyn2011}.

\subsubsection{Equilibrium solution}
We can immediately find a trivial solution of the divergence-free condition.
When the probabilistic flow in a steady state vanishes, the divergence-free condition is satisfied:
\begin{equation}
{\bf J}^{\rm (eq)}({\bf x}) = {\bf 0},
\end{equation}
where the superscript has been changed from ``{\rm ss}" to ``{\rm eq}".
This steady state is in particular called  the equilibrium state because no current exists.
Then the force becomes 
\begin{equation}
{\bf A}({\bf x}) = - {\rm grad}E({\bf x}).
\end{equation}
This is the trivial solution in terms of our purpose.
Indeed, the resulting force does not violate the DBC.
This fact can be confirmed through a formulation of the transition probability.
To violate the DBC, we add an additional term to the trivial force.
Below we show several nontrivial solutions, which can indeed violate the DBC.

\subsubsection{Vector-field solutions}
Let us move on from the case without current.
We demand the probabilistic flow in a steady state to take the following form:
\begin{equation}
{\bf J}^{\rm (ss)}({\bf x}) = \gamma{\bf B}({\bf x})P^{\rm (ss)}({\bf x}),
\end{equation}
where $\gamma$ is a control parameter on the degree of violation of the DBC and will be detailed below.
Owing to the divergence-free condition, the vector field ${\bf B}({\bf x})$ must satisfy
\begin{equation}
0 = \gamma\sum_{k=1}^N \left( \frac{\partial [{\bf B}]_k}{\partial [{\bf x}]_k}
- \frac{1}{T}[{\bf B}]_k\frac{\partial E}{\partial [{\bf x}]_k}
\right)P^{\rm (ss)}({\bf x}) ,
\end{equation}
where we have used the fact that the distribution function is set to be $P^{\rm (ss)}({\bf x}) \propto \exp(-E({\bf x})/T)$ and the subscript for the bracket denotes the element of the vector.
A trivial solution is given by 
\begin{equation}
[{\bf B}({\bf x})]_k = \exp(E({\bf x})/T - F/T).
\end{equation}
Then the probabilistic flow becomes a constant current as
\begin{equation}
{\bf J}^{\rm (ss)}({\bf x}) = \gamma {\bf 1}.
\end{equation}
Here ${\bf 1}$ denotes a vector with all elements unity.
We obtain the resulting force as
\begin{equation}
{\bf A}({\bf x}) = - {\rm grad}E({\bf x}) + \gamma \exp(E({\bf x})/T - F/T){\bf 1}.
\end{equation}
For practical use of this force, we may rescale $\gamma \exp(-F/T)$ as $\gamma$.
However, we have to be careful about the instability in using  the exponential term, in particular in the low-temperature region.

Let us find an alternative to the above obtained force to remove the instability of the exponential term.
We set the vector field as
\begin{equation}
[{\bf B}({\bf x})]_k = \left(\frac{\partial E}{\partial [{\bf x}]_{k-1}} - \frac{\partial E}{\partial [{\bf x}]_{k+1}}\right),\label{nont1}
\end{equation}
where $[{\bf x}]_{N+1} = [{\bf x}]_1$ and $[{\bf x}]_0 = [{\bf x}]_N$.
Then the divergence-free condition holds.
The resulting force is given as
\begin{equation}
[{\bf A}({\bf x})]_k = -\frac{\partial E}{\partial [{\bf x}]_k} + \gamma \left(\frac{\partial E}{\partial [{\bf x}]_{k-1}} - \frac{\partial E}{\partial [{\bf x}]_{k+1}}\right).\label{DF1}
\end{equation}
This force drives the system while satisfying the divergence-free condition but neglecting the DBC.
In this sense, the resulting force is analogous to the case of using the Suwa-Todo method in the context of the MCMC method \cite{Suwa2010}.
This type of  solution is not unique.
We may permute the index $j=1,2,\dots,N$ as $j = P(1),P(2),\dots,P(N)$, where $P(\cdot)$ denotes a permutation of $N$ indices.
To reach the best performance, we may pursue an optimal permutation to realize the fastest convergence.
This is beyond our scope in the present study.
Let us further find any other nontrivial solutions.

\subsection{Divergence-free condition for a duplicate system}
As in the case of the SDBC \cite{Turitsyn2011}, let us duplicate the system.
Then we may neglect the DBC for each system but hold the BC for the composite system.
The analogy to the SDBC leads to the duplicate Fokker-Planck equation 
\begin{equation}
\frac{\partial P({\bf x}_1,{\bf x}_2,t)}{\partial t} = - \sum_{i=1}^2{\rm div}_i{\bf J}_i({\bf x}_1,{\bf x}_2,t)\label{DF2},
\end{equation}
where ${\bf x}_i$ is the $N$-dimensional vector for each system denoted by $i$ and ${\bf J}_i({\bf x}_1,{\bf x}_2,t)$ is the probabilistic flow defined as
\begin{equation}
{\bf J}_i({\bf x}_1,{\bf x}_2,t) = \left({\bf A}_i({\bf x}_1,{\bf x}_2) - T {\rm grad}_i\right)P({\bf x}_1,{\bf x}_2,t).
\end{equation}
The gradient indexed by $i$ is taken for each system.
Again we impose the divergence-free condition.
The divergence-free condition for the duplicate system can be recast as
\begin{equation}
0 = - \sum_{i=1}^2{\rm div}_i{\bf J}^{\rm (ss)}_i({\bf x}_1,{\bf x}_2)\label{BCLan2},
\end{equation}
where
\begin{equation}
{\bf J}^{\rm (ss)}_i({\bf x}_1,{\bf x}_2) = \left({\bf A}_i({\bf x}_1,{\bf x}_2) - T {\rm grad}_i\right)P^{\rm (ss)}({\bf x}_1)P^{\rm (ss)}({\bf x}_2).
\end{equation}
We impose the steady state of the duplicate system as $P^{\rm (ss)}({\bf x}_1,{\bf x}_2) = P^{\rm (ss)}({\bf x}_1) P^{\rm (ss)}({\bf x}_2)$.
We can find a nontrivial solution by considering the mixture of the forces on the duplicate system as
\begin{eqnarray}
{\bf A}_1({\bf x}_1,{\bf x}_2) &=& - {\rm grad}_{1}E({\bf x}_1) + \gamma  {\rm grad}_{2}E({\bf x}_2)\label{DF21} ,\\
{\bf A}_2({\bf x}_1,{\bf x}_2) &=& - {\rm grad}_{2}E({\bf x}_2) - \gamma  {\rm grad}_{1}E({\bf x}_1)\label{DF22}.
\end{eqnarray}
Then the probabilistic flow in the steady state is given as
\begin{eqnarray}
{\bf J}^{\rm (ss)}_1({\bf x}_1,{\bf x}_2) &=& \gamma {\rm grad}_{2}E({\bf x}_2)P^{\rm (ss)}({\bf x}_1)P^{\rm (ss)}({\bf x}_2) ,\\
{\bf J}^{\rm (ss)}_2({\bf x}_1,{\bf x}_2) &=& - \gamma {\rm grad}_{1}E({\bf x}_1)P^{\rm (ss)}({\bf x}_1)P^{\rm (ss)}({\bf x}_2).
\end{eqnarray}
One can confirm that the divergence of the above probabilistic flows can vanish immediately because the summation becomes zero.

One may recall the exchange Monte Carlo simulation, in which a multiple system with different temperatures is driven simultaneously, while  each realization is exchanged several times.
The technique exhibits extremely outstanding performance with fast convergence to the equilibrium state.
Naturally, we may expect the existence of an analogous Langevin dynamics to the exchange Monte Carlo simulation.
We here mention this  fascinating problem as a subject of a future study, which will be published elsewhere.

\subsection{Path probability and detailed balance condition}
As shown in the previous subsection, we find nontrivial solutions satisfying the analogous condition to the BC (\ref{BCLan}) and (\ref{BCLan2}).
We here confirm the violation of the DBC in these solutions.
Let us calculate the transition probability during an infinitesimal time interval, $[t,t+dt]$.
We obtain
\begin{eqnarray}
&&L^{\gamma}({\bf x}[t+dt]|{\bf x}[t])\propto \exp\left\{-\frac{1}{4T}(\dot{{\bf x}} - {\bf A})^2dt -\frac{1}{2}{\rm div} {\bf A} dt\right\},\label{transition}
\end{eqnarray}
where ${\bf x}[t]$ is the location at time $t$.
For the duplicated system, a similar computation yields the transition probability.
We here omit the arguments of the quantities on the right-hand side to simplify the expressions.
We use the midpoint prescription, $\bar{{\bf x}} = ({\bf x}[t+dt]+{\bf x}[t])/2$; we take the partial derivative with respect to the location.
The ratio of the transition probability between the forward and backward processes confirms violation of the DBC owing to the existence of the probabilistic flow as follows:
\begin{eqnarray}
& &\frac{L^{\gamma}({\bf x}[t+dt]|{\bf x}[t])}{L^{\gamma}({\bf x}[t]|{\bf x}[t+dt])} 
= \exp\left( \frac{\gamma}{T} \dot{{\bf x}}^{\rm T} {\bf B} dt - \frac{1}{T} \dot{{\bf x}}^{\rm T} {\rm grad} E(\bar{{\bf x}}) dt\right).
\end{eqnarray}
The first term in the exponential function corresponds to the negative of the time-derivative housekeeping heat $\dot{Q}_{\rm hk}$, while the second term can be regarded as the negative of the time-derivative excess heat $\dot{Q}_{\rm ex}$, i.e., \cite{Hatano2001,Sughiyama2011}
\begin{eqnarray}
\dot{Q}_{\rm hk} &=& - \gamma\dot{{\bf x}}^{\rm T} {\bf B},\\
\dot{Q}_{\rm ex} &=& \dot{{\bf x}}^{\rm T} {\rm grad}E(\bar{{\bf x}}).
\end{eqnarray}
The existence of the housekeeping heat signals violation of the DBC and, in this sense, $\gamma$, which is the coefficient in the vector field ${\bf B}$, controls the degree of violation of the DBC.
Also in the case of the duplicate system, we can compute the ratio of the transition probabilities for forward and backward processes as
\begin{eqnarray} \nonumber
& & \prod_{i=1,2}\frac{L^{\gamma}_i({\bf x}_i[t+dt]|{\bf x}_i[t])}{L^{\gamma}_i({\bf x}_i[t]|{\bf x}_i[t+dt])} 
= \exp\left( \frac{\gamma}{T} (\dot{{\bf x}_1}^{\rm T}{\rm grad}_2 E(\bar{{\bf x}_2}) -\dot{{\bf x}_2}^{\rm T}{\rm grad}_1 E(\bar{{\bf x}_1})) dt\right) 
\\
& & \quad \times \prod_{i=1,2}\exp\left( - \frac{1}{T} \dot{{\bf x}}^{\rm T} {\rm grad}_i E(\bar{{\bf x}_i}) dt\right).
\end{eqnarray}
Then the time derivatives of the housekeeping heat and excess heat are given as
\begin{eqnarray}
\dot{Q}_{\rm hk} &=& \gamma\left(\dot{{\bf x}_2}^{\rm T}{\rm grad}_1 E(\bar{{\bf x}_2}) -\dot{{\bf x}_1}^{\rm T}{\rm grad}_2 E(\bar{{\bf x}_1})\right),\\
\dot{Q}_{\rm ex} &=& \sum_{i=1,2}\dot{{\bf x}_i}^{\rm T} {\rm grad}_iE(\bar{{\bf x}_i}).
\end{eqnarray}
In the duplicate system, we also confirm the violation of the DBC from the existence of the housekeeping heat.
Let us below evaluate the performance of using the artificial force violating the DBC by the above nontrivial solutions.

\section{Numerical tests}
We implement the above artificial force to introduce the probabilistic flow violating the DBC.
To confirm that the artificial force accelerates the relaxation to the steady state, we demonstrate  simple numerical tests below.
We here restrict ourselves to the case of the additional force to the duplicate system.
Let us consider the system with the following double-valley potential energy:
\begin{equation}
U(x) = - \frac{1}{2}x^2 + \frac{1}{4}x^4.
\end{equation}
The minima of the potential are located at $x=-1$ and $1$.
The equilibrium point is $x=0$; that is, the average over independent runs converges to $\langle x[t] \rangle \equiv \sum_{k=1}^{N_{\rm sam}}x^{(k)}[t]/N_{\rm sam}$ in $t \to \infty$, where $x[t]$ denotes the realization at the time $t$.
We simulate the independent $N_{\rm sam} = 1000$ runs.
The initial condition is set to be at $x^{(k)}[0] = 1$.
To achieve the equilibrium point, the particles must climb up the local maximum and reach the other minimum of the potential energy.
Therefore, it takes a relatively long time to reach the equilibrium point.
Then we introduce the additional force to accelerate the stochastic dynamics and duplicate the system.
The time evolution of the Langevin equation is evaluated by using the ordinary method known as the Heun scheme \cite{Kloeden2011}.
We set the infinitesimal time as $d t = 0.0001$.
We take an average over independent  $N_{\rm sam}=1000$ runs, while taking the time average during $\Delta t = 0.1$.

\begin{figure}[tb]
\begin{center}
\includegraphics[width=0.6\textwidth]{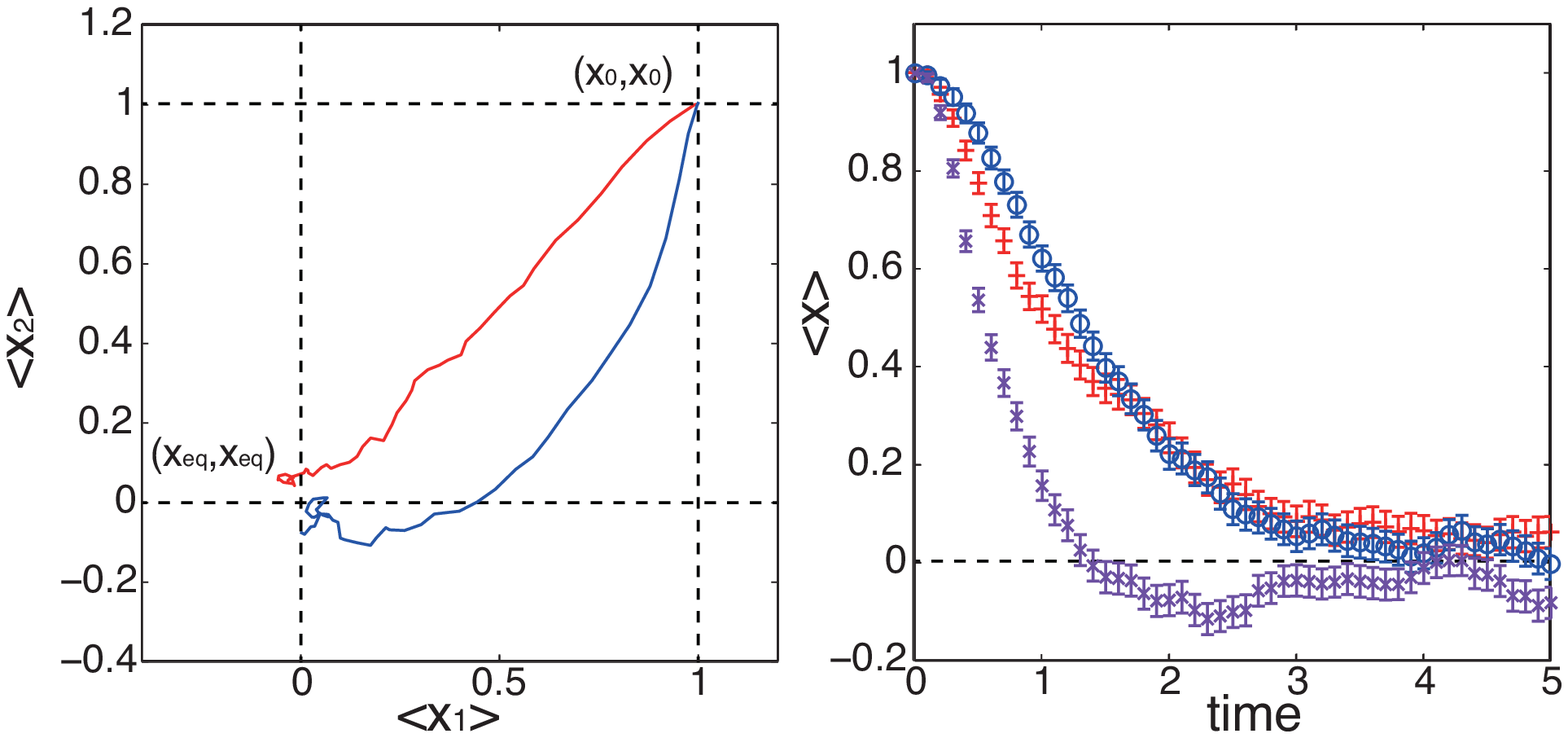}
\includegraphics[width=0.6\textwidth]{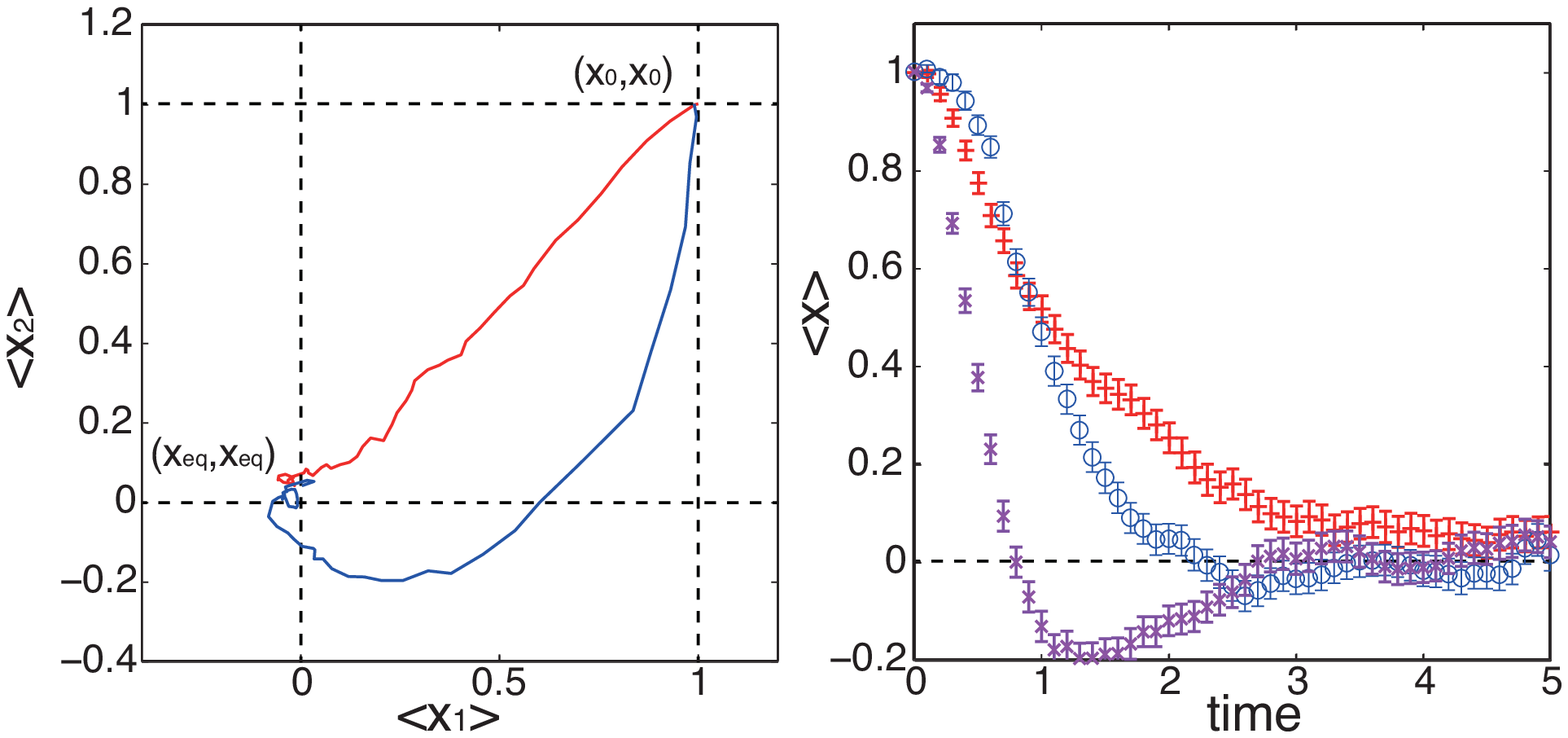}
\includegraphics[width=0.6\textwidth]{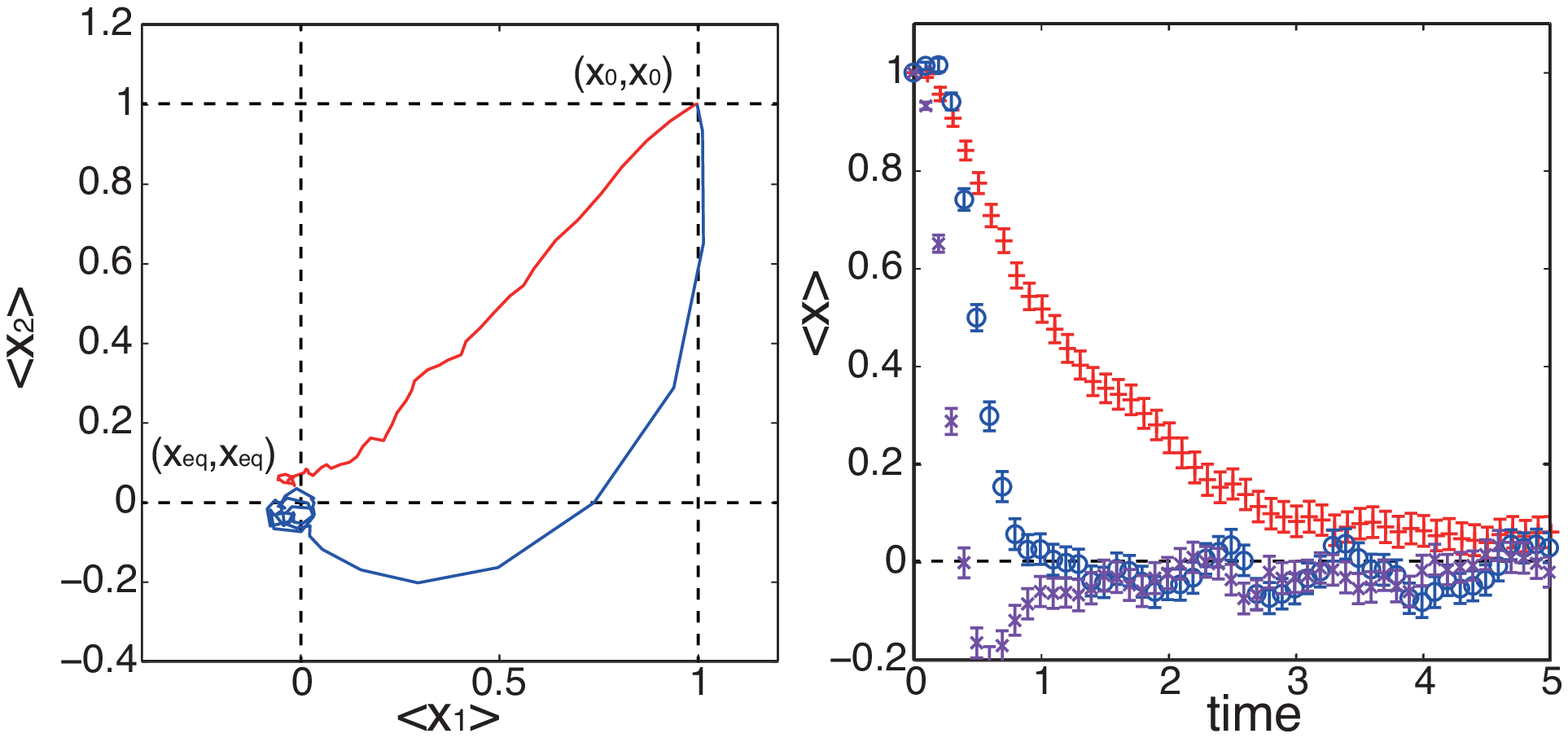}
\includegraphics[width=0.6\textwidth]{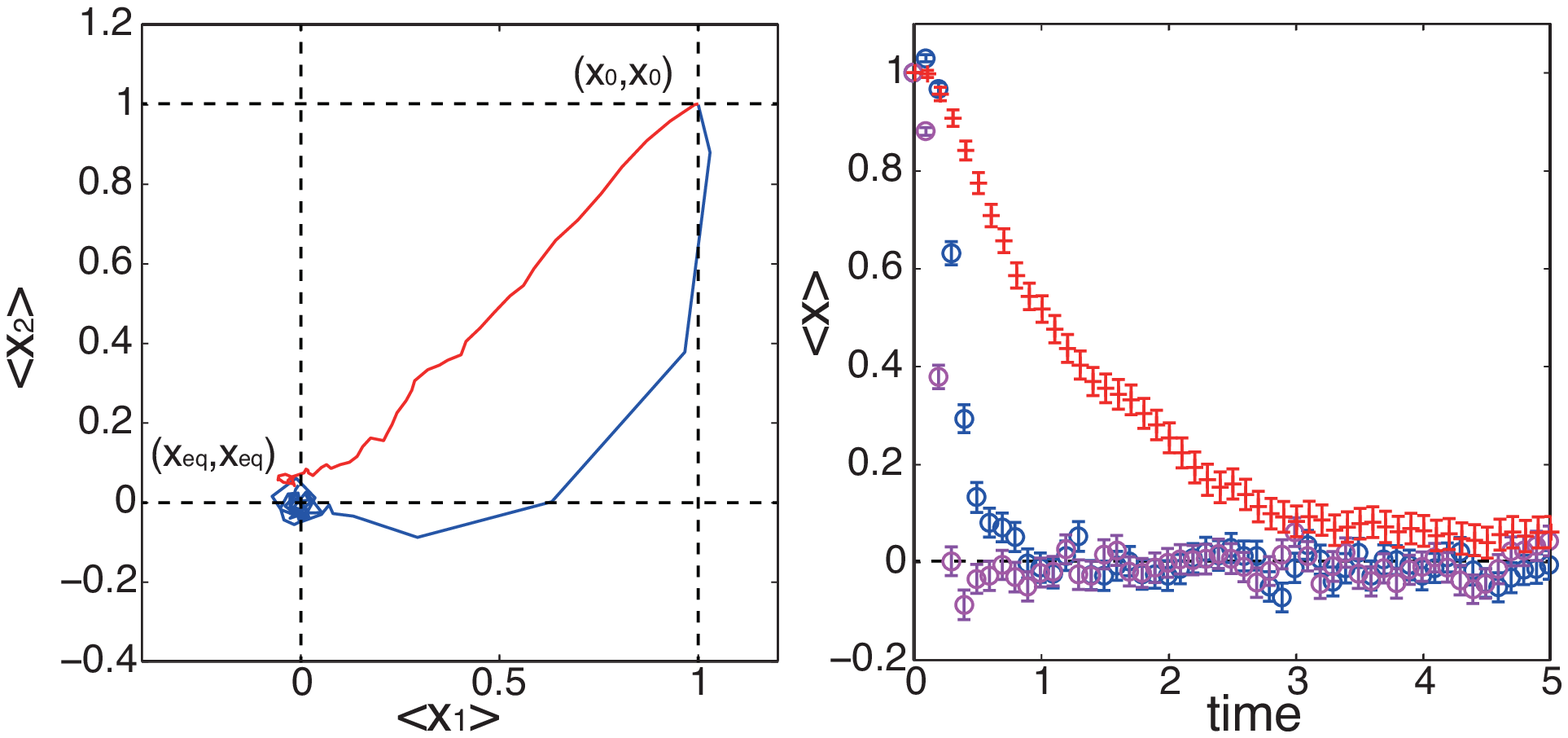}
\end{center}
\caption{Comparison between the cases with and without additional force.
From top to bottom, the values of $\gamma$ take $1$, $2$, $5$, and $10$.
In the left panels, the red orbit denotes the case without additional force.
The blue ones represent the case with additional force on the plane of $x_1$ and $x_2$, which is the location of a particle in the duplicate system.
The right panels show the average of the location over independent runs over $N_{\rm sam}=1000$.
Red plots are for the case without additional force and the blue and purple ones stand for each  average of the location of the duplicate system  with additional force.
}
\label{fig1}
\end{figure}

\begin{figure}[tb]
\begin{center}
\includegraphics[width=0.4\textwidth]{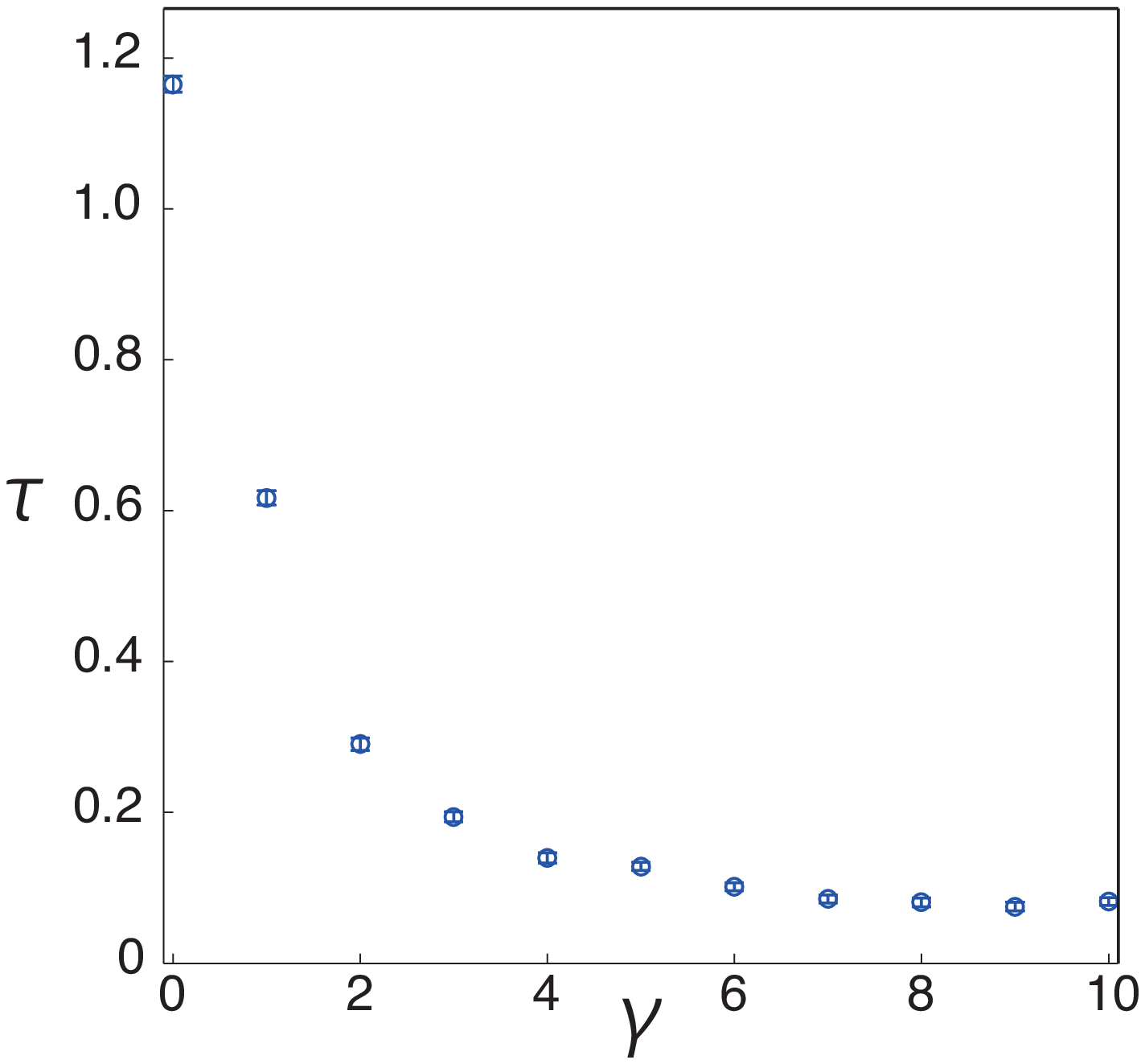}
\end{center}
\caption{Integrated correlation time.
The horizontal axis denotes the value of $\gamma$ and the vertical one describes the average of the integrated correlation time.
The statistical error falls within the size of the data points.
}
\label{fig2}
\end{figure}

Figure \ref{fig1} shows a comparison between the cases without and with additional force.
We describe the orbit on the plane of the locations of the duplicate system.
The averaged locations of the duplicate system are expressed as $\langle x_1 \rangle$ and $\langle x_2 \rangle$. 
The case without additional force traces the line between $(0,0)$ and $(1,1)$ because any difference does not exist in the duplicate system.
In contrast, the case with additional force describes a different path from the line.
The particles seem to wrap around the swelled area of the potential energy in the plane of $(x_1,x_2)$.
For weak $\gamma$, the relaxation to the steady state is not necessarily faster.
However, an increase of the value of $\gamma$ leads to  faster convergence to the steady state  than that for the case without additional force.
This is one of the benefits in violating the DBC, as shown in  previous studies in terms of the Suwa-Todo method and the SDBC.
In addition, we observe the integrated correlation time in the steady state.
The integrated correlation time is defined as the collection of the correlation times:
\begin{equation}
\tau_i = \sum_{t=1}^{\infty} \tau^{O}_{i,t} = \frac{\langle O_iO_{i+t} \rangle - \langle O_i\rangle \langle O_{i+t} \rangle}{\langle O^2 \rangle - \langle O \rangle^2}.
\end{equation}
In the steady state, the dependence of the correlation time on $i$ is lost.
In numerical computation, we cannot sum the correlation time $\tau^{O}_{i,t}$ up to an infinite number.
Thus the dependence on $i$ remains.
We then take an average over $i$ to avoid this dependence.
In this computation, we use the location as the physical observable. 
The results are shown in Figure \ref{fig2}.
We confirm the decrease of the integrated correlation time.
In addition, the integrated correlation time reaches a lower limit against any further increase of the value of $\gamma$.
The decrease of the correlation time gives a reduction of the necessary number to efficiently compute the averaged value from the sampling and thus leads to an increase in the efficacy of the sampling.
This beneficial point is the same as the one found in the Suwa-Todo method and the SDBC.

In addition, the present authors demonstrated a further application of our method to a classical spin model, the XY model, which has a critical slowing down in the low-temperature region \cite{Ohzeki2013}.
The proposed method exhibited  outstanding performance, removing the critical slowing down of the XY model, which often hampers efficient computation in the steady state.

The remaining problem is the mathematical and theoretical understanding of the faster convergence to the steady state and the reduction of the correlation time owing to violation of the DBC.
Fortunately, the former property can be explained by considering the eigenvalues of the operator form of the Fokker-Planck equation and transition-rate matrix in the master equation.
In the next section, we demonstrate the property of the operator form of the Fokker-Planck equation in  cases without and with additional forces and show the shift of the eigenvalues such that it promotes accelerating the relaxation to the steady state.

\section{Mathematical assurance: analysis of eigenvalues}
To clarify   the origin of the acceleration to the steady state by the additional force, we rewrite the Fokker-Planck equation in operator form \cite{Risken1996}.
This expression can help us to find a mathematical understanding of the acceleration of the relaxation to the steady state in our technique.
\subsection{Operator form of the Fokker-Planck equation}
First we rescale the instantaneous distribution as 
\begin{equation}
\tilde{P}({\bf x},t) = \frac{P({\bf x},t)}{\sqrt{P^{\rm (ss)}({\bf x})}}.
\end{equation}
Then the Fokker-Planck equation in the equilibrium system can be rewritten as
\begin{equation}
\frac{\partial \tilde{P}({\bf x},t)}{\partial t} =  \sum_{k=1}^N \left( \frac{1}{4T} \left( \frac{\partial E}{\partial [{\bf x}]_k}\right)^2 + \frac{1}{2}\frac{\partial^2 E}{\partial [{\bf x}]_k^2} - T \frac{\partial^2}{\partial [{\bf x}]_k^2} \right) \tilde{P}({\bf x},t).
\end{equation}
Let us here define the following operators as
\begin{eqnarray}
\left[{\bf a}\right]_k = \frac{1}{2\sqrt{T}}\frac{\partial E}{\partial [{\bf x}]_k}  + \sqrt{T}\frac{\partial}{\partial [{\bf x}]_k},\\
\left[{\bf a}^{\dagger}\right]_k = \frac{1}{2\sqrt{T}}\frac{\partial E}{\partial [{\bf x}]_k}  - \sqrt{T}\frac{\partial}{\partial [{\bf x}]_k},
\end{eqnarray}
where ${\bf a}^\dagger$ denotes the adjoint of ${\bf a}$.
These operators satisfy the following commutation relation:
\begin{equation}
\left[ \left[{\bf a}^{\dagger}\right]_k, \left[{\bf a}\right]_l \right] = \frac{\partial^2 E}{\partial [{\bf x}]_k \partial [{\bf x}]_l}
\end{equation}
and trivial ones as $\left[ \left[{\bf a}\right]_k, \left[{\bf a}\right]_l \right] = 0$ and $\left[ \left[{\bf a}^{\dagger}\right]_k, \left[{\bf a}^{\dagger}\right]_l \right] = 0$.
The Fokker-Planck equation is simply expressed in terms of these operators as 
\begin{equation}
\frac{\partial \tilde{P}({\bf x},t)}{\partial t} = S({\bf a}^{\dagger},{\bf a})\tilde{P}({\bf x},t),
\end{equation}
where 
\begin{eqnarray}\nonumber
S({\bf a}^{\dagger},{\bf a}) &=& - {{\bf a}^{\dagger}}^{\rm T} {\bf a}.
\end{eqnarray}
Here we use the transpose of the vectors denoted by $(\cdot)^{\rm T}$.
The operator on the right-hand side in the Fokker-Planck equation is Hermitian.
The square root of the Gibbs-Boltzmann distribution can be an eigenfunction of this operator with zero eigenvalue.
By using the same operator expression, we can rewrite the Fokker-Planck equation for the nontrivial solution satisfying the divergence-free condition (\ref{DF1}).
The additional force modifies the Fokker-Planck equation into 
\begin{eqnarray}\nonumber
\frac{\partial \tilde{P}({\bf x},t)}{\partial t} &=& S({\bf a}^{\dagger},{\bf a})\tilde{P}({\bf x},t) \\
& & + \gamma \sum_{k=1}^N \left( \frac{1}{2T}[{\bf B}]_k \frac{\partial E}{\partial [{\bf x}]_k} - \frac{\partial [{\bf B}]_k}{\partial [{\bf x}]_k} -  [{\bf B}]_k\frac{\partial }{\partial [{\bf x}]_k} \right)\tilde{P}({\bf x},t).
\end{eqnarray}
From this observation, we find that the effect of the additional force in the Fokker-Planck equation can be given by $\gamma {{\bf a}^{\dagger}}^{\rm T}{\bf B}({\bf x})\tilde{P}({\bf x},t)/\sqrt{T}$.
The nontrivial solution of the vector field (\ref{nont1}) can be expressed by the operators as  
\begin{equation}
[{\bf B}({\bf x})]_k = \sqrt{T}\left([{\bf a}^{\dagger}]_{k-1} + [{\bf a}]_{k-1} - [{\bf a}^{\dagger}]_{k+1} - [{\bf a}]_{k+1}\right).
\end{equation}
Thus we reach
\begin{equation}
\frac{\partial \tilde{P}({\bf x},t)}{\partial t} = W({\bf a}^{\dagger},{\bf a})\tilde{P}({\bf x},t)\label{FP2},
\end{equation}
where $\tilde{P}({\bf x},t) = P({\bf x},t)/\sqrt{P^{\rm (ss)}({\bf x})}$ and 
\begin{eqnarray}\nonumber
W({\bf a}^{\dagger},{\bf a}) &=& S({\bf a}^{\dagger},{\bf a}) - \gamma\sum_{k=1}^N
[{\bf a}^{\dagger}]_{k}(\left[{\bf a}\right]_{k-1} - \left[{\bf a}\right]_{k+1}).
\end{eqnarray}
We find the anti-Hermitian part on the right-hand side of the Fokker-Planck equation (\ref{FP2}).
The anti-Hermitian part does not change the steady state because the eigenfunction of the original Fokker-Planck equation is that for the annihilation operator ${\bf a}$. 
Thus we expect that the existence of the anti-Hermitian part of the operators on the right-hand side is closely related to the acceleration of the relaxation.
In addition, the trivial solution with the exponentially divergent quantity also yields the anti-Hermitian part in  operator form.

Furthermore, the above observation is supported from the Fokker-Planck equation for the divergence-free duplicate system (\ref{DF2}).
The Fokker-Planck equation for the duplicate system with nontrivial additional force can be rewritten as
\begin{eqnarray}\nonumber
\frac{\partial \tilde{P}({\bf x}_1,{\bf x}_2,t)}{\partial t} &=& \sum_{i=1}^2 S({\bf a}_i^{\dagger},{\bf a}_i)\tilde{P}({\bf x}_1,{\bf x}_2,t)\\
& & - \gamma \sum_{k=1}^N \left(\frac{\partial E}{\partial [{\bf x}_2]_k} \frac{\partial }{\partial [{\bf x}_1]_k} - \frac{\partial E}{\partial [{\bf x}_1]_k} \frac{\partial }{\partial [{\bf x}_2]_k} \right)\tilde{P}({\bf x}_1,{\bf x}_2,t),
\end{eqnarray}
where $\tilde{P}({\bf x}_1,{\bf x}_2,t) = P({\bf x}_1,{\bf x}_2,t)/\sqrt{P^{\rm (ss)}({\bf x}_1,{\bf x}_2)}$.
Here we define the operators ${\bf a}_i$ and ${\bf a}_i^{\dagger}$ for each system.
The commutation relation can be replaced with
\begin{equation}
\left[ \left[{\bf a}^{\dagger}_i\right]_k, \left[{\bf a}_j\right]_l \right] = \delta_{ij}\frac{\partial^2 E}{\partial [{\bf x}_i]_k \partial [{\bf x}_i]_l}
\end{equation}
and trivial ones as $\left[ \left[{\bf a}_i\right]_k, \left[{\bf a}_j\right]_l \right] = 0$ and $\left[ \left[{\bf a}^{\dagger}_i\right]_k, \left[{\bf a}^{\dagger}_j\right]_l \right] = 0$.
We here use the Kronecker delta $\delta_{ij}$, which takes unity when $i=j$ and otherwise vanishes.
Thus we reach
\begin{equation}
\frac{\partial \tilde{P}({\bf x}_1,{\bf x}_2,t)}{\partial t} = W(\{{\bf a}_i^{\dagger}\},\{{\bf a}_i\})\tilde{P}({\bf x}_1,{\bf x}_2,t)\label{FP22},
\end{equation}
where 
\begin{equation}
W(\{{\bf a}_i^{\dagger}\},\{{\bf a}_i\}) = \sum_{i=1}^2 S({\bf a}_i^{\dagger}, {\bf a}_i) - \gamma\left({{\bf a}_2^{\dagger}}^{\rm T}{\bf a}_1 - {{\bf a}_1^{\dagger}}^{\rm T}{\bf a}_2\right).
\end{equation}
The artificial force satisfying the divergence-free condition involves the anti-Hermitian part also in the case for the duplicate system.
From this common property, the anti-Hermitian part of the operators in the Fokker-Planck equation is expected to accelerate the relaxation.
In addition, when we violate the DBC in the master equation, namely, $W({\bf x}|{\bf y}) \neq W({\bf y}|{\bf x})$ in the expression in Equation (\ref{PV}), we find that the transition matrix takes the asymmetric form.
The key point comes from the asymmetry of the transition matrix in the master equation or operators in the Fokker-Planck equation.
The relaxation time is characterized by the gap between the first and second largest eigenvalues of the transition matrix and operators.
Below let us analyze the eigenvalues for asymmetric operators by identifying both of them for simplicity.
Then we prove that the gap of the cases with the additional force is always opened in comparison to that without it.

\subsection{Shift of the second eigenvalues}
\begin{figure}[tb]
\begin{center}
\includegraphics[width=0.8\textwidth]{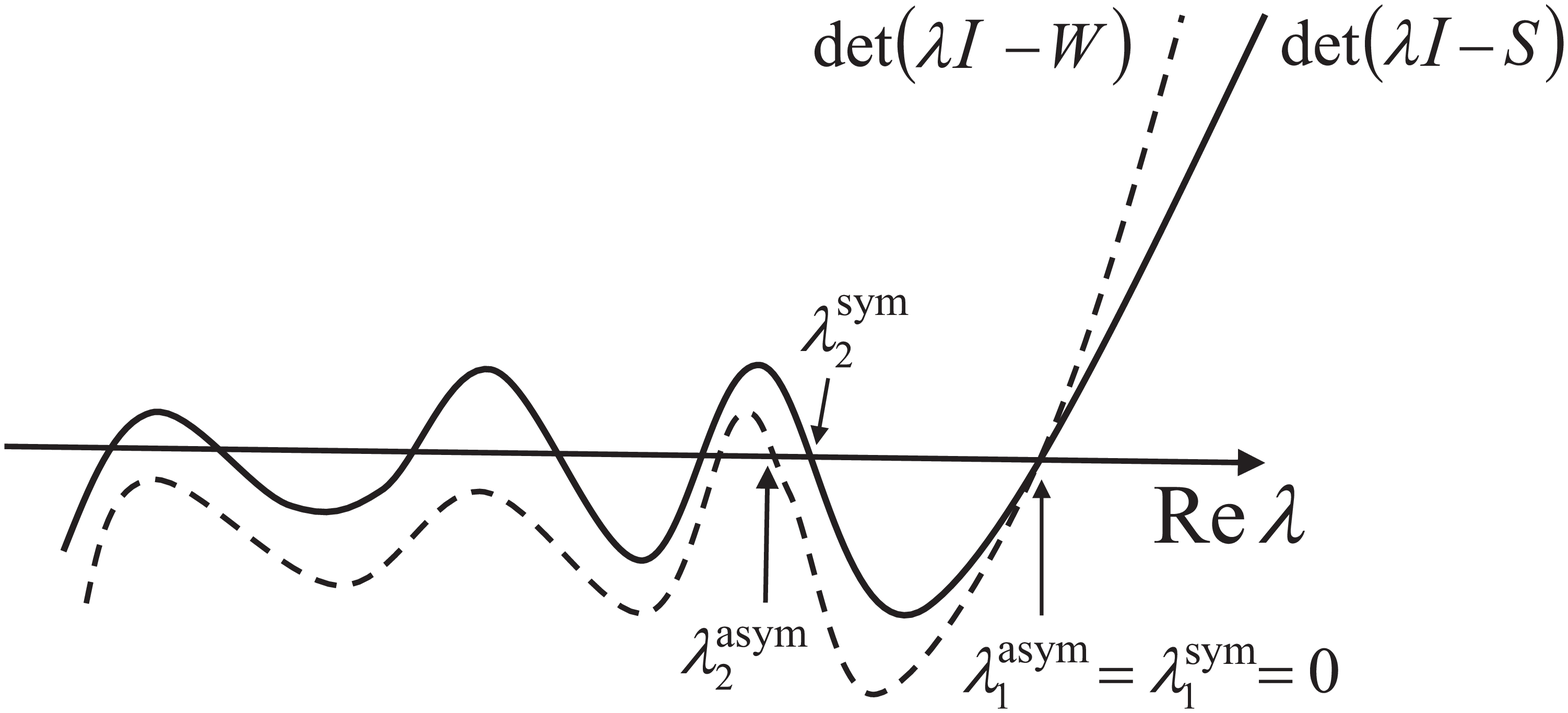}
\end{center}
\caption{Real parts of $\det\left(\lambda I-W({\bf a}^\dagger, {\bf a})\right)$ and $\det\left(\lambda I-S({\bf a}^\dagger, {\bf a})\right)$.
Inequality (\ref{second}) claims $\det\left(\lambda I-W({\bf a}^\dagger, {\bf a})\right)\le\det\left(\lambda I-S({\bf a}^\dagger, {\bf a})\right)$ for $\lambda_2^{\rm (sym)}<\lambda<0$.}
\label{fig3}
\end{figure}

Let us consider the eigenvalues for the asymmetric matrix and operators.
We restrict ourselves to the case of the nontrivial solution of the vector field in the single system.
The following analysis can be straightforwardly generalized to the   duplicate system case.
The trivial eigenfunction for the operator $W({\bf a}^\dagger, {\bf a})$ is $\sqrt{P^{\rm (ss)}({\bf x})}$ with zero eigenvalue that is the largest.  
Thus relaxation to steady state $P^{\rm (ss)}({\bf x})$ is assured \cite{Risken1996}.
The relaxation speed is dominated by the gap between the first and the second largest eigenvalues.  
As seen above, introduction of the artificial force induces the anti-Hermitian part of the operator $W({\bf a}^\dagger, {\bf a})$, while the Hermitian part is fixed.  
We here utilize a mathematical tool from linear algebra known as the Ostrowski-Taussky inequality \cite{Horn2012}
\begin{equation}
\left|\det A\right| \ge \det\frac{A+A^\dagger}{2},\label{OT}
\end{equation}
where $A$ is an arbitrary operator with $(A+A^\dagger)/2$ being positive definite.
We can show that the anti-Hermitian part opens the gap larger by applying the Ostrowski-Taussky inequality to the characteristic polynomial $\det\left(\lambda I-W({\bf a}^\dagger, {\bf a})\right)$, where $I$ is an identity operator.
Since the largest eigenvalues for both $W({\bf a}^\dagger, {\bf a})$ and $\left(W({\bf a}^\dagger, {\bf a})+W^\dagger({\bf a}^\dagger, {\bf a})\right)/2$  are  $\lambda_1^{\rm (asym)}=\lambda_1^{\rm (sym)}=0$, the Ostrowski-Taussky inequality yields 
\begin{equation}
\left|\det\left(\lambda I-W({\bf a}^\dagger, {\bf a})\right)\right|\ge\det\left(\lambda I-S({\bf a}^\dagger, {\bf a})\right)\label{largest}
\end{equation}
for arbitrary real $\lambda>0$. 
We show a schematic picture of the behavior of two characteristic functions in Figure \ref{fig3}.
The root of both of the characteristic functions is located at $\lambda=0$.
In addition, Inequality (\ref{largest}) ensures that the slope of the characteristic function of the asymmetric operator at $\lambda=0$ is steeper than that of the corresponding symmetric part. 
To prove the fact that the second largest eigenvalue, the nearest root to the origin, is located at the further point by induction of the anti-Hermitian operator, let us consider $\det\left(\lambda I-W({\bf a}^\dagger, {\bf a})\right)/\lambda$, which has the same roots as those of the characteristic function except for $\lambda=0$.
We can prove the lemma that the Hermitian and anti-Hermitian operators $\tilde{S}_\lambda$ and $\tilde{G}$ exists satisfying
\begin{equation}
\frac{\det\left(\lambda I-W({\bf a}^\dagger, {\bf a})\right)}{\lambda}=\det\left(\tilde{S}_\lambda({\bf a}^\dagger, {\bf a})-\tilde{G}({\bf a}^\dagger, {\bf a})\right)
\end{equation}
for arbitrary real $\lambda$ using the appropriate unitary transformation.  
Since the unitary transformation does not change the symmetries and determinants of the operators, we reach 
\begin{equation}
\det\left(\tilde{S}_\lambda({\bf a}^\dagger, {\bf a})\right)=\frac{\det\left(\lambda I-S({\bf a}^\dagger, {\bf a})\right)}{\lambda}.
\end{equation}
For $\lambda$ satisfying $\lambda_2^{\rm (sym)}<\lambda<0$, where $\lambda_2^{\rm (sym)}$ denotes the second largest eigenvalue for the symmetric operator, $\tilde{S}_\lambda({\bf a}^\dagger, {\bf a})$ is positive definite.  
Resorting the Ostrowski-Taussky inequality, we find 
\begin{equation}
\left|\frac{\det\left(\lambda I-W({\bf a}^\dagger, {\bf a})\right)}{\lambda}\right|\ge\frac{\det\left(\lambda I-S({\bf a}^\dagger, {\bf a})\right)}{\lambda}.\label{second}
\end{equation}
This inequality shows that the second largest eigenvalue of the asymmetric operator is further from the origin than that of the corresponding symmetric operator.
All of the above ingredients shows that 
\begin{equation}
{\rm Re }\lambda_2^{\rm (asym)}\le \lambda_2^{\rm (sym)}.
\end{equation}
Therefore, the acceleration of convergence by the introduction of asymmetry in the transition dynamics is guaranteed in terms of the shift of eigenvalues.  

\section{Conclusion}
We investigated several solutions of nontrivial forces to accelerate convergence to the desired distribution in Langevin systems.
The solutions are roughly categorized into two types.
The first is the addition of an exponentially divergent force to the gradient of the potential, i.e., natural force.
However, this type of force involves numerical instabilities such as overflow.
In contrast, the second solution, which takes advantage of rotation of the vector field, is free from such numerical instabilities.
The latter solution can be simplified by employing the idea in the duplicated system inspired by the SDBC.
Our proposed method has been indeed confirmed by numerical tests to accelerate convergence.
This aspect of violation of the DBC can give mathematical assurance in terms of the shift of eigenvalues for asymmetric operators.
In numerical experiments, we also confirm that our method shortens the correlation time compared to  that in the ordinary implementation of  Langevin dynamics.
Previous studies on   using MCMC method without the DBC also revealed the reduction of the correlation time, and we may thus regard this property as a beneficial aspect of the DBC violation. 
This advantage might be explained by some mathematical property on the violation of the DBC, but we have not reached any conclusive result in this regard.
We hope that a future study will clarify the reduction of the correlation time for the case violating the DBC.

In the present study, we utilize a mathematical tool in linear algebra to prove the acceleration of convergence to the steady state.
However, recent developments in nonequilibrium statistical mechanics can explain two of the beneficial points of the violation of the DBC.
The acceleration of convergence and reduction of the correlation time rely on the change of frequency of state transitions.
Let us consider the case with local minima of potential energy.
The existence of local minima can be a bottleneck in the convergence toward the equilibrium state.
This is because the typical events are to be trapped in the local minima.
As shown in our present study,  violation of the DBC can help the particles escape from local minima and accelerates the relaxation to the steady state.
This implies that the particles pass through the modified path from the typical transitions in the equilibrium case. 
The modification of the path probability between different microscopic states by additional forces has been recently discussed in the context of  biased sampling in nonequilibrium statistical mechanics \cite{Nemoto2011,Sughiyama2013}.
The detailed analysis on the effect of the additional force will be reported elsewhere from the viewpoint of biased sampling and its optimality by considering a kind of  variational principle.
Such a viewpoint will give a deeper understanding on the faster convergence to the steady state resulting from the violation of the DBC and fundamental aspects on nonequilibrium statistical mechanics through  concrete instances near the equilibrium system.

\section*{Acknowledgments}
One of the authors (M.O.) is grateful for fruitful discussions with M. Kikuchi, H. Touchette, and K. Hukushima.
This work was supported by MEXT in Japan, KAKENHI No. 24740263 and No. 25120008, and the Kayamori Foundation of Informational Science Advancement.
\section*{References}
\bibliography{Paper_ver2}
\end{document}